\documentclass[twocolumn, RNAAS]{aastex702}
\usepackage{xspace}
\usepackage{enumitem}

\newcommand{\salt}{{SALT}\xspace}

\shorttitle{Optical Spectroscopy of TeV-emitting BL Lac candidates}
\shortauthors{Metzger, Hickox, \& Thorstensen}

\begin{document}

\title{Optical Spectroscopy of TeV-emitting BL Lac Candidates}

\author[0000-0003-3585-3356]{Cassidy Metzger}
\affiliation{Department of Physics \& Astronomy Dartmouth College, 6127 Wilder Laboratory, Hanover, NH 03755, USA}
\email{cassidy.m.metzger.gr@dartmouth.edu}
\correspondingauthor{C. Metzger}
\email{cassidy.m.metzger.gr@dartmouth.edu}
\author[0000-0003-1468-9526]{Ryan C. Hickox}
\email{Ryan.C.Hickox@dartmouth.edu}
\affiliation{Department of Physics \& Astronomy Dartmouth College, 6127 Wilder Laboratory, Hanover, NH 03755, USA}
\author[0000-0002-4964-4144 ]{John R. Thorstensen}
\email{John.R.Thorstensen@dartmouth.edu}
\affiliation{Department of Physics \& Astronomy Dartmouth College, 6127 Wilder Laboratory, Hanover, NH 03755, USA}
\begin{abstract}
TeV-emitting BL Lacertae (BL Lac) blazars have important implications for the study of jet phenomenology, particle acceleration, and ultra high-energy cosmic ray production. They also allow for indirect studies of the extragalactic background light, cosmic magnetic fields, and axion-like particles. The upcoming Cherenkov Telescope Array (CTA), a ground-based gamma-ray observatory, will expand the observed TeV-emitting BL Lac population and enhance the aforementioned fields of study. However, reliable redshifts are crucial for planning and interpreting observations with CTA and only about 50\% of gamma-ray BL Lacs have spectroscopic redshifts. We performed medium resolution optical spectroscopy of 16 TeV-emitting BL Lac candidates with the \salt and MDM telescopes. We measured spectroscopic redshifts for the full sample, ranging from 0.059 to 0.4, including 5 new spectroscopic redshift measurements. 
\end{abstract}


\section{Introduction} \label{sec:intro}
Blazars are active galactic nuclei that exhibit relativistic outflows along the observer's line of sight \citep{Urry_1995}. Their spectral energy distributions peak at low energies from synchrotron radiation and again at high energies from inverse-Compton scattering. In rare cases, the high-energy peak exceeds 10 TeV, challenging standard synchrotron self-Compton models. \citep{Costamante_2018}. 
Only 56 BL Lacs have been detected beyond 0.1 TeV (\texttt{TeVCAT}\footnote{ \href{http://tevcat2.uchicago.edu}{http://tevcat2.uchicago.edu}}; \citealp{Wakely_2008}), but the Cherenkov Telescope Array (CTA) is expected to reveal several hundred new TeV-peaked blazars, enabling population-level
statistics for the first time \citep{CTA_2019}. 


High precision redshifts are crucial for planning and interpreting observations with CTA,  but BL Lacs' near featureless, continuum-dominated spectra make redshifts difficult to obtain. To address this, we retrieved optical spectra for 16 objects that are proposed as TeV-emitting BL Lac candidates by \citet{Chang_2019}, \citet{Arsioli_2015}, and \citet{Metzger2025}. 


\section{Observations and analysis}\label{sec:data}
We completed 14 observations with the Robert Stobie Spectrograph (RSS; \citealp{Burgh2003}) mounted on the 11-m Southern African Large Telescope (\salt) between November 2025 and January 2026, and two additional observations in April 2026 with the Ohio State Multi-object Spectrometer (OSMOS; \citealp{Martini2011}) on the 2.4 m Hiltner reflector at the MDM Observatory. For \salt targets, we obtained three 750-second exposures with the PG0900 grating in Long Slit Spectroscopy mode, using a 1.5$''$ slit for 5.67\AA~resolution from $4500$ to $7500$~\AA. For the two MDM targets, we obtained three 1200-second exposures with the "blue" disperser and a $1''.1$ "inner" slit, yielding a $3.1$~\AA~resolution (FWHM) from $3980$ to $6860$~\AA. 

\salt spectra were reduced following standard IRAF protocol \citep{Tody1086} for cross-talk, bias, gain, and flat-field correction. Source and background spectra from MDM were extracted using the protocol of \citet{Horne_1986}. For the wavelength calibrations we took Hg, Ne, and Xe lamp spectra at zenith and then adjusted those solutions slightly to minimize the velocities of the airglow features in the background. 
To derive redshifts, we searched each Gaussian-smoothed spectrum for faint stellar absorption features of the host galaxy, cross-checked candidates against other features at the same redshift, and fit each with a Gaussian profile. Monte Carlo resampling was used to compute uncertainty on the Gaussian center, with uncertainty from the quadrature sum of redshift variation and wavelength calibration error. 
We took the average of the Gaussian centers to compute the redshift of each source and summed the changes to the redshift in quadrature with the error on the wavelength calibration to determine the redshift uncertainty. 

\section{Results}{\label{sec:results}}
We resolved at least two absorption features for all 16 objects and recovered spectroscopic redshifts for the entire sample. A literature review revealed that 11 targets had previously reported spectroscopic redshifts, 3 had photometric redshifts, and 2 had no prior redshift measurements. Where multiple literature values existed, we chose the one best matching our result as the fiducial. Our results are summarized in 
Table \ref{table:1}.

We found good agreement with all prior spectroscopic redshifts except for J2221-524. \citet{Jones_2009} reported $z = 0.24$, but flagged this measurement to be low-quality. \citet{Shaw_2013} and \citet{DAmmando_2024} found the spectrum to be featureless. We resolved the CaII H and K doublet and retrieved a redshift of 0.4, consistent with the photometric bounds from \citet{Rau_2012} and \citet{Arsioli_2015}. 

We also note disparities for J0956-099 and J0123-231. For J0956-099, prior spectroscopic redshifts of 0.16 \citep{Grazian_2002} and 0.14 \citep{Guzzo_2009} exist; our result ($z = 0.158^{+0.0011}_{-0.00249}$) best matches \citet{Guzzo_2009}'s findings. For J0123-231, prior values of 0.4 \citep{Bauer_2000} and 0.2 \citep{Jones_2009} exist; our result ($z = 0.397^{+0.00352}_{-0.00103}$) is in closest agreement with \citet{Bauer_2000}.

We present novel spectroscopic redshift measurements for J0001-419, J2322-492, J0647-515, J0143-587, and J0826-640. Our results for J0647-515 and J0826-640 fall within $\pm 0.05$ of the photometric redshifts obtained by \citet{Chang_2019}, while J2322-492 deviates by $+0.14$. J0001-419, J0647-515, and J0143-587 had been observed spectroscopically before without visible absorption or emission lines \citep{Marais_2024, Pena_2021, Landoni_2015}. 

In summary, we present 5 new spectroscopic redshifts, a revised redshift for J2221-524, resolution of disparities for J0956-099 and J0123-231, and confirm agreement with prior spectroscopic redshifts for 8 sources.

\begin{deluxetable*}{cccccccc} 
\tabletypesize{\scriptsize}

\tablecaption{List of observed sources and their corresponding redshifts.}
\tablehead{\colhead{THC} & \colhead{RA} & \colhead{DEC} & \colhead{$z_{\text{new}}$}& \colhead{$z_{\text{old, fiducial}}$}& \colhead{Redshift type}& \colhead{Literature source} & \colhead{$m_G$}
}
\startdata
J0917-037 & 09:17:35               & -03:43:34               & 0.310$^{+0.0011}_{-0.0025}$ & 0.31                        & S         & AC16        & 19.08           \\
J0336-037 & 03:36:24               & -03:47:39               & 0.162$^{+0.0031}_{-0.0020}$                                           &$ 0.16\pm 0.00018  $                    & S    & J09         & 18.46              \\
J0956-099 & 09:56:50               & -09:57:37               & 0.158$^{+0.0011}_{-0.0025}$                                           & $0.16\pm0.00077$                                & S    & Gu02      & 19.06                 \\
J0123-231 & 01:23:38               & -23:10:59               & 0.397$^{+0.0035}_{-0.0010}$                                           & 0.40                 & S     & B00    &  18.47              \\
J0001-419 & 00:01:33               & -41:55:26               & 0.137$^{+0.0020}_{-0.0013}$                                           & --                    & --      & --   & 18.36                   \\
J2322-492 & 23:22:54               & -49:16:30               & 0.240$^{+0.0025}_{-0.0071}$                                           & 0.38                   & P  & C19      & 16.77                  \\
J0647-515 & 06:47:10               & -51:35:48               & 0.186$^{+0.0012}_{-0.0038}$                                           & 0.22        & P    & C19         & 18.38           \\
J2221-524 & 22:21:29               & -52:25:26               & 0.400$^{+0.0010}_{-0.0011}$                                           & $0.24 $           & S           & J09 & 16.70             \\
J0209-524 & 02:09:22               & -52:29:24               & 0.214$^{+0.0082}_{-0.0011}$                                           & $0.21\pm 0.00020 $                        & S   & Go21   & 16.39                      \\
J0156-530 & 01:56:58               & -53:01:59               & 0.304$^{+0.0013}_{-0.0011}$                                          & $0.30\pm0.00040$                  & S     & Go21   &   18.41                \\
J0543-555 & 05:43:57               & -55:32:06               & 0.271$^{+0.0010}_{-0.014}$                                           & 0.27              & S  & P14      & 15.80                    \\
J0325-565 & 03:25:24               & -56:35:46               & 0.059$^{+0.0014}_{-0.0014}$                                           & $0.060\pm0.00015  $              & S          & J09        & 17.51      \\
J0244-583 & 02:44:40               & -58:19:55               & 0.260$^{+0.0013}_{-0.0011}$                                           & $0.27 \pm 0.00015 $             & S      & J09         &16.76       \\
J0143-587 & 01:43:48               & -58:45:50               & 0.372$^{+0.0013}_{-0.0011}$                                           & --                   & --     &       --      & 16.74        \\
J0826-640 & 08:26:28               & -64:04:15               & 0.289$^{+0.0011}_{-0.0012}$                                           & 0.24                    & P    & C19    & 17.42             \\
J0352-685 & 03:52:58               & -68:31:16               & 0.086$^{+0.0014}_{-0.0059}$                                          & 0.087                & S     & F98      & 17.52            
\enddata

\vspace{0.1 cm}
\hspace{0.1 cm} {Sources are ordered by descending declination. Columns: THC designation; RA and Dec (J2000); redshift derived in this work; best-matching literature redshift and type (P = photometric, S = spectroscopic) with citation \citep{Alvarez_2016, Jones_2009, Guzzo_2009, Bauer_2000, Chang_2019, Arsioli_2015, Pita_2014, Fischer_1998}}; Gaia g-band magnitude (Gaia Collaboration et al. 2023).
\label{table:1}

\end{deluxetable*}

\facilities{\salt, MDM}

\software{astropy \citep{2013A&A...558A..33A,2018AJ....156..123A,2022ApJ...935..167A}, IRAF \citep{Tody1086, Tody_1993}}.

\begin{acknowledgments}
Observations were obtained with SALT under program 2025-2-SCI-031 (PI: Hickox). We thank Justin Anderson at MDM Observatory for assistance and Professor Chris Lintott for his helpful review. C.M. acknowledges support from a Dartmouth Fellowship.

\end{acknowledgments}

\bibliography{references}
\bibliographystyle{aasjournal}
\end{document}